\documentclass[conference]{IEEEtran}
\IEEEoverridecommandlockouts
\usepackage{cite}
\usepackage{amsmath,amssymb,amsfonts}
\usepackage{algorithmic}
\usepackage[numbers]{natbib}
\usepackage{graphicx}
\usepackage{subfigure}
\usepackage{textcomp}
\usepackage{xcolor}
\def\BibTeX{{\rm B\kern-.05em{\sc i\kern-.025em b}\kern-.08em
    T\kern-.1667em\lower.7ex\hbox{E}\kern-.125emX}}
    
\begin{document}

\title{GAEA: Experiences and Lessons Learned from a Country-Scale Environmental Digital Twin}

\author{\IEEEauthorblockN{Andreas Kamilaris\IEEEauthorrefmark{1}\IEEEauthorrefmark{2}, Chirag Padubidri\IEEEauthorrefmark{1}, Asfa Jamil\IEEEauthorrefmark{1}, Arslan Amin\IEEEauthorrefmark{1}, Indrajit Kalita\IEEEauthorrefmark{1}, Jyoti Harti\IEEEauthorrefmark{1}, \\ Savvas Karatsiolis\IEEEauthorrefmark{1} and Aytac Guley\IEEEauthorrefmark{1}}
\IEEEauthorblockA{\IEEEauthorrefmark{1}\textit{SuPerWorld Research Group, CYENS Center of Excellence, Nicosia, Cyprus} \\
\IEEEauthorblockA{\IEEEauthorrefmark{2}\textit{Pervasive Systems Group, University of Twente, Enschede, The Netherlands} \\
* Corresponding Author's email: a.kamilaris@cyens.org.cy}}}

\maketitle

\begin{abstract}
This paper describes the experiences and lessons learned after the deployment of a country-scale environmental digital twin on the island of Cyprus for three years. This digital twin, called GAEA, contains 27 environmental geospatial services and is suitable for urban planners, policymakers, farmers, property owners, real-estate and forestry professionals, as well as insurance companies and banks that have properties in their portfolio. This paper demonstrates the power, potential, current and future challenges of geospatial analytics and environmental digital twins on a large scale.
\end{abstract}


\section{Introduction}
Modern pervasive computing technologies have unlocked an unprecedented capacity to create high-fidelity digital twins of our natural and built environments \cite{mihai2022digital, nativi2021digital}. Advances in cloud computing provided the scalable computational backbone needed to process big data, while sophisticated machine and deep learning models offer the analytical power to translate raw data from the field and remote sensing into predictive analytics. This technological synergy enables the construction of dynamic, data-driven virtual replicas of the real world that are not confined to the controlled industrial settings of a few square kilometers, but can encompass vast and complex geographical areas \cite{ketzler2020digital}. These environmental digital twins (EDTs) move beyond static models to become living systems that simulate real-world processes, forecast future scenarios, and provide a critical decision-support platform for a wide range of stakeholders, from urban planners to financial institutions \cite{blair2021digital, nativi2021digital}. Worth-mentioning examples of EDTs include EarthMap \cite{morales2023earth} (visualization, processing, and analysis of global land and climate data), GeoEngine \cite{verma2022geoengine} (geospatial analytics to simulate real-world conditions and provide actionable insights for cross-sector decision-making) and Virtual Singapore \cite{ignatius2019virtual} (a large-scale 3D model of the city and a collaborative data platform).

The realization of EDTs on a large scale (i.e., a country size, thousands of square kilometers) is based on the integration of a diverse spectrum of technologies \cite{mihai2022digital}. Satellite imagery and earth observation data from programs like Copernicus and Landsat provide a macro-scale, historical view of environmental changes, while field sensing and near real-time data streams from ground-based sensors monitor specific locations and provide a micro-scale, frequently updated view of the world. Geospatial analytics then fuses these multi-modal data streams, layering in topographic, demographic, and infrastructural information to create a semantically rich and spatially accurate digital counterpart. This holistic approach allows for the modeling of complex inter-dependencies, such as simulating the impact of a new development on local biodiversity or assessing flood risk for a portfolio of properties.

The true potential of EDTs and the insights they generate is only realized through the active participation and engagement of their beneficiaries and end users \cite{blair2021digital} (see Section \ref{Stakeholders}). For urban planners and policymakers, such insights facilitate evidence-based zoning and climate resilience planning. Farmers can leverage insights on soil and microclimate conditions for precision agriculture, while property owners and real-estate professionals can assess long-term sustainability and value risks. Crucially, for insurance companies and banks, the information on hazard, vulnerability and exposure at high spatial granularity allows them to characterize the physical risks of their portfolio, especially in a changing climate, where disasters are expected more frequently and to a larger extent \cite{van2006impacts}. Therefore, fostering an ecosystem of shared knowledge—through EDT platforms and insight-rich visualization tools is crucial to ensuring that these powerful digital twins drive informed action across society and industry.

The contribution of this paper is to record and communicate the experiences and lessons learned after the implementation, deployment, and evaluation of an EDT for the geographical area of the Republic of Cyprus, called GAEA \cite{jamil2024gaea}. The authors aspire that this paper can serve as a useful handbook for EDT planners and developers in the future, to better align their ambitions and designs with the expectations of potential beneficiaries and local markets.

\section{GAEA Environmental Digital Twin}
\begin{figure*}[htbp]
    \subfigure[]{\includegraphics[width=0.5\textwidth]{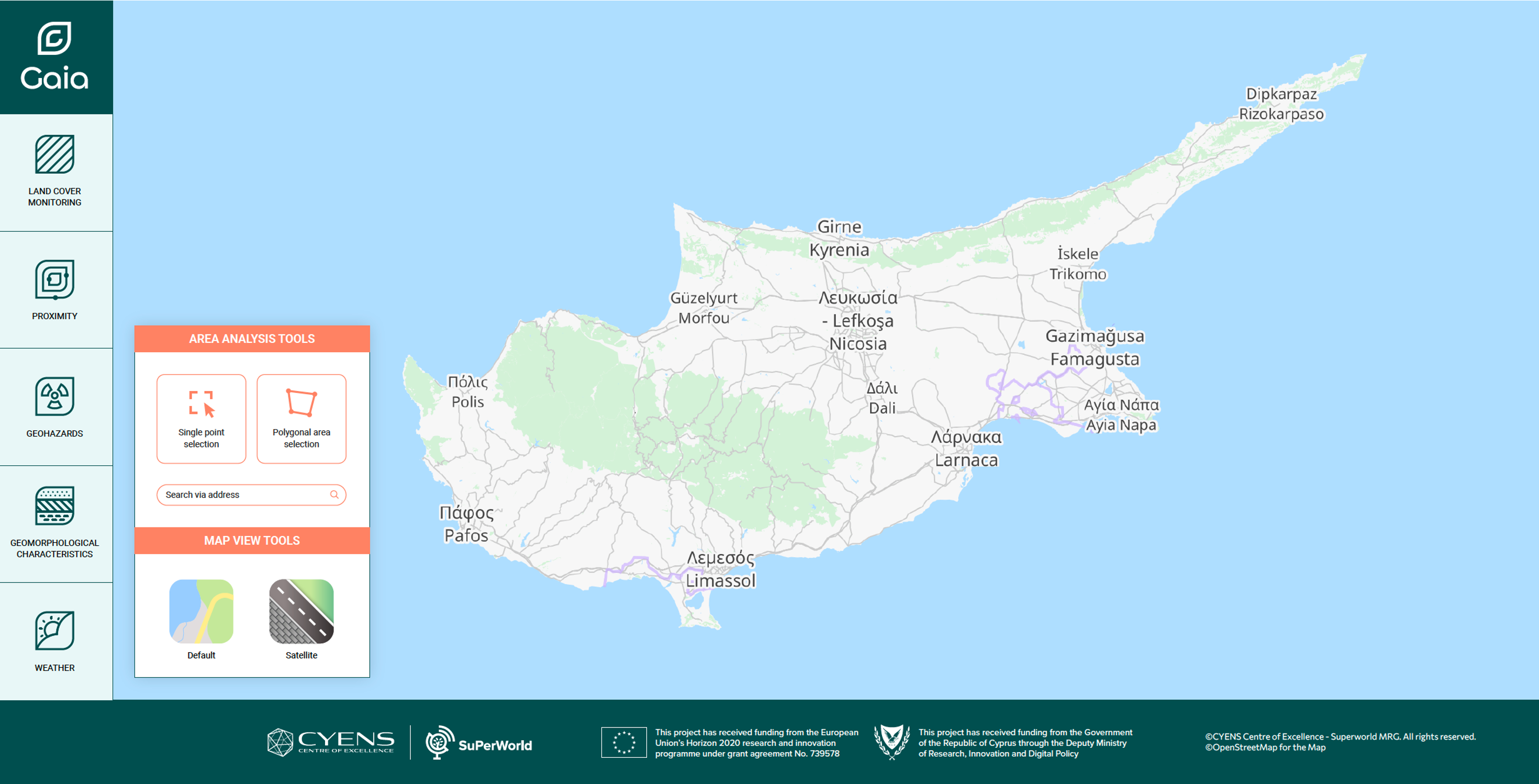}} 
    \subfigure[]{\includegraphics[width=0.5\textwidth]{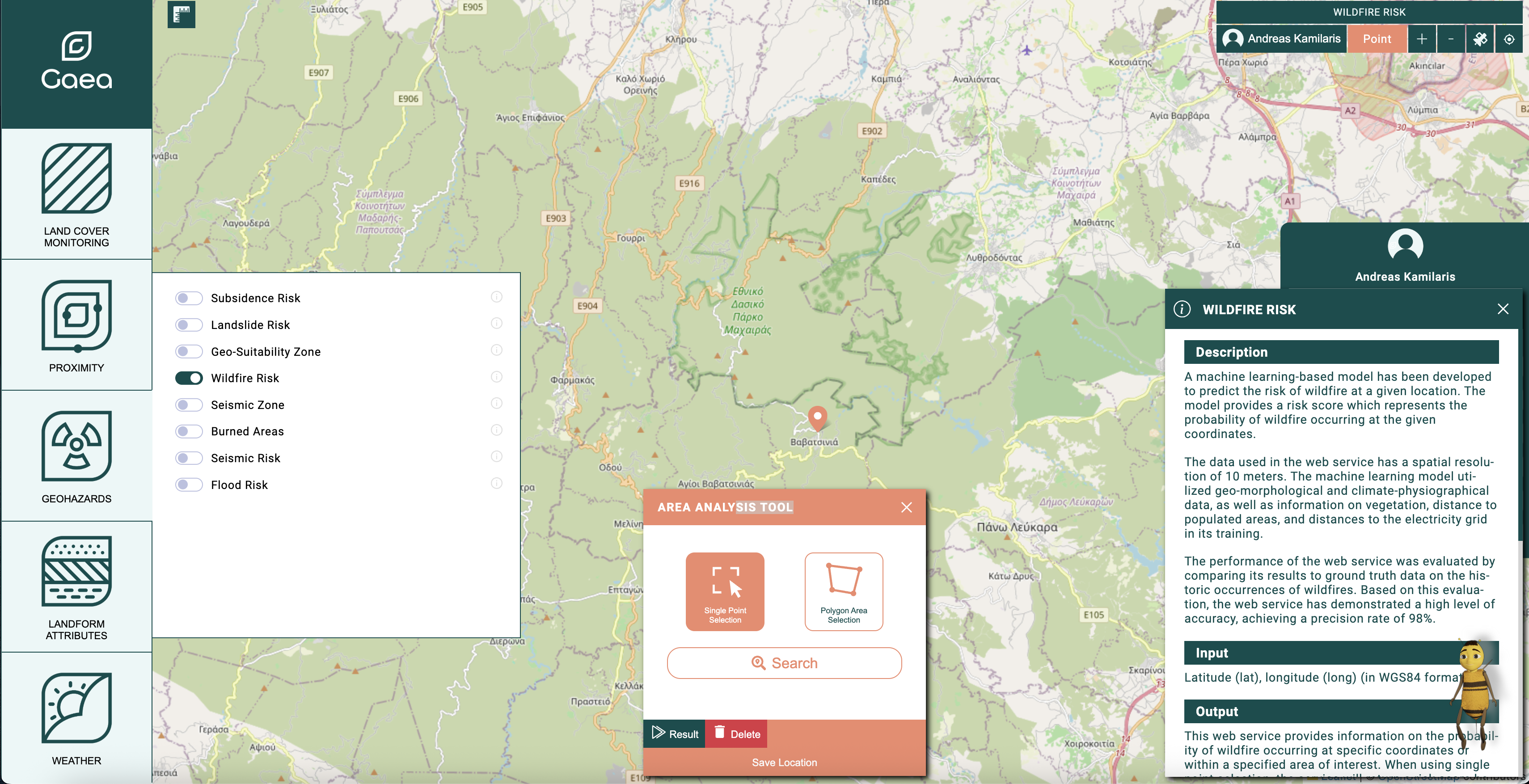}} 
    \subfigure[]{\includegraphics[width=0.5\textwidth]{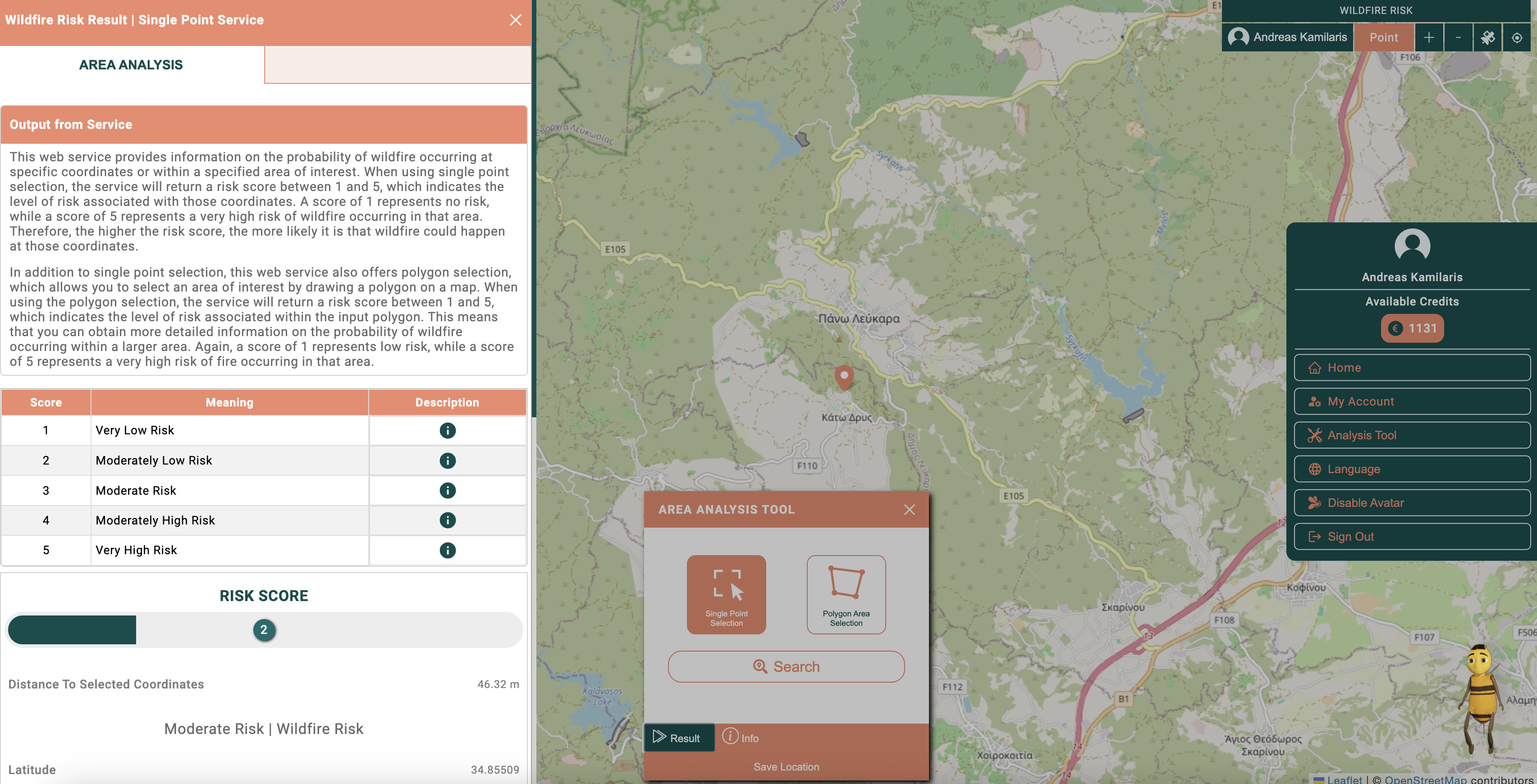}}
    \subfigure[]{\includegraphics[width=0.5\textwidth]{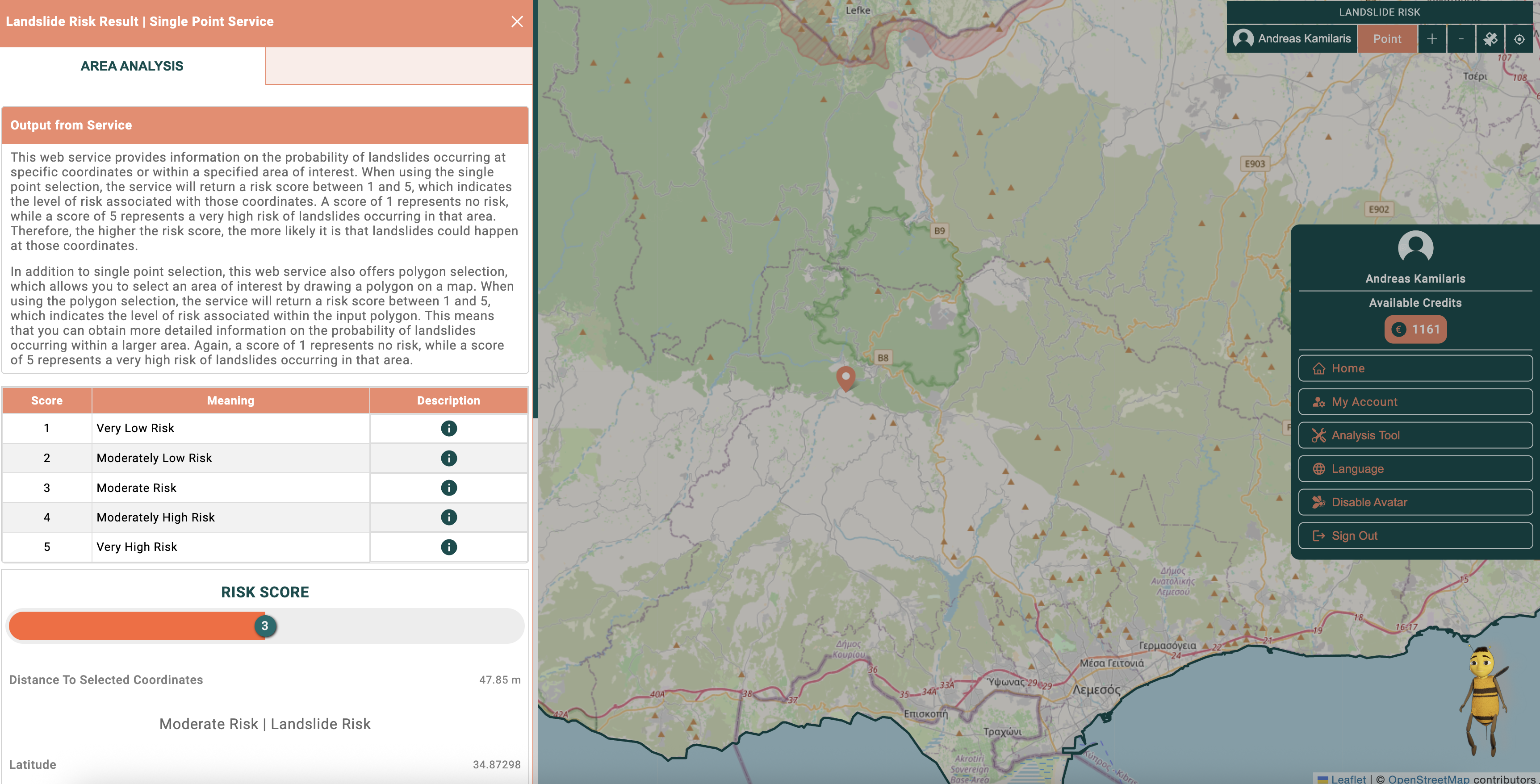}}
\caption{Snapshots from GAEA front-end. (a) Main dashboard showing group of services on the left vertical menu. (b) Selection of the wildfire risk estimation service. (c) Results from the wildfire risk service, for a mountainous location with moderately row risk. (d) Results from the landslide risk service, for a rural area with moderate risk.}
\label{fig1}
\end{figure*}

The GAEA geoanalytical tool\footnote{GAEA. http://gaea.cyens.org.cy}, developed by the SuPerWorld\footnote{SuPerWorld research group. https://superworld.cyens.org.cy/} research group of the CYENS Center of Excellence\footnote{CYENS CoE. https://cyens.org.cy} in 2023, is an AI-empowered interactive online tool that offers 27 geoanalytical services\footnote{GAEA Services. https://superworld.cyens.org.cy/product1.html}, covering the whole geographical area of the Republic of Cyprus. Combining advanced earth observation technologies, AI, and geospatial analysis techniques, GAEA provides understanding of the impact of environmental risks and climatic change on any location at the geographical area of the Republic of Cyprus. Via GAEA, users can compare different locations and identify trends and patterns both locally and regionally, in regards to the physical environment and geospatial trends. GAEA incorporates land cover mapping and land use change over time, providing a wealth of information for informed decision-making. Gaea offers a range of services, including the detection of swimming pools and buildings, vegetation and burnt areas nearby the property, natural risk estimation (land subsidence, landslides, wildfires, flooding, earthquakes), slope and aspect, geology, precipitation,
elevation, land use, proximity to roads, sea, blue-flag beaches, amenities, electricity network, and the Natura 2000. This allows for a comprehensive assessment of the suitability and potential risks of an area and is useful for farmers, real estate agents, and property owners. Screenshots of the front-end of GAEA are provided in Figure \ref{fig1}. Technical details about GAEA are available in Jamil et al. \cite{jamil2024gaea}.

\section{Interaction with Stakeholders}
\label{Stakeholders}
During these three years, we interacted with numerous local stakeholders, each with different requirements, providing feedback based on their specific needs. In this section, we describe our interactions with some key stakeholders.

\subsubsection{Banks and insurance companies}
Banks and insurance companies have a portfolio of assets they need to insure by incorporating physical risks related to these properties. In Cyprus, key physical risks include wildfires, flooding, landslides, earthquakes, and land subsidence/displacement. Three local banks interacted with us, two of them purchased our hazard estimation services. These stakeholders need to know both the hazard probability for each property, plus the consequent financial exposure, quantified as estimated repair costs. GAEA supports the classification of hazards into five classes, described in Table \ref{tab1}. The quantification of these hazard classes into damage functions was conducted in collaboration with local insurance partners, leveraging historical claims data correlated with specific disaster events. Among all perils, wildfires constitute the primary concern for these institutions. The risk is sufficiently acute that it has prompted a cessation of insurance coverage for properties in mountainous regions, particularly ones of wooden construction. Furthermore, risk managers from all three participating banks explicitly inquired about the projected impact of climate change on these risk profiles, with emphasis on wildfires and flooding.

\begin{table*}[htbp]
\caption{Hazard risk classification.}
\begin{center}
\begin{tabular}{|c|p{16.5cm}|}
\hline
\textbf{Class} & \textbf{\textit{Explanation}}  \\
\hline
1 & Minimal probability of occurrence; highly unlikely event. No historical incidents of disaster at that location and/or the local geological/weather/other parameters do not suggest a risk for a disaster to occur.  \\
\hline
2 &  Low probability of occurrence; Moderate-low intensity. Very few historical incidents of disaster and/or the local geological/weather/other parameters slightly suggest a risk for a disaster to occur.\\
\hline
3 &  Moderate probability of occurrence; Possible mild events with moderate intensity. Few historical incidents of disaster and/or the local geological/weather/other parameters suggest a certain risk for a disaster to occur. \\
\hline
4 &  High probability; Considerable occurrence likelihood; Moderate-high intensity; Minor damages expected. Many historical incidents of disaster and/or the local geological/weather/other parameters suggest a considerable risk for a disaster to occur. \\
\hline
5 &  Very high probability; Significant occurrence likelihood; Very high intensity expecting certain damages. A large number of historical incidents of disaster and/or the local geological/weather/other parameters suggest a high risk for a disaster to occur. \\
\hline
\end{tabular}
\label{tab1}
\end{center}
\end{table*}

\subsubsection{Real estate agents}
\label{realestate}
Agents of real estate properties were mostly interested in the distance of each property to the nearest amenities (i.e., supermarkets, hospitals, pharmacies, schools) and to the beach, considering that Cyprus is a touristic island surrounded by sea. The agents also asked for distances to the nearest road (even dirty roads) and to the overall green (i.e., vegetation) of the surroundings. This geospatial intelligence was not easy to obtain from other sources, and they found GAEA valuable in this regard. In addition, one agent motivated us to consider this added value to improve existing automated valuation models (AVMs). He provided us with historic records of purchase prices of 30,000 properties, and we trained ML models to predict their sale prices both with and without the additional insights provided by GAEA. Our results showed that additional geospatial insights increase the performance of AVMs by 3-5\% \cite{gravier2024leveraging}.

\subsubsection{Property owners}
GAEA was visited by ca. 500 owners of properties around Cyprus. This user group was mainly interested in the physical risks of their properties (see Figure \ref{fig4}). The communication of high-risk assessments to stakeholders was carefully designed to avoid undue alarm and instead foster constructive engagement (see Table \ref{tab1}). When surveyed, a subset of users whose properties were identified as being in high-risk zones reported that this objective data heightened their awareness of climate change impacts. Furthermore, they expressed an increased intention to advocate for bettter climate mitigation and adaptation measures from their local communities and governmental bodies.

\subsubsection{Municipalities}
Municipalities were particularly interested in vegetation statistics within their geographical areas, with respect to the number of trees and the types of popular trees within their regions. They wanted to learn the relationship between greenery and construction and how this changed during the last 10 years. Certain municipalities (Larnaka and Limassol) were interested in the number of swimming pools and their total volume, as this relates to their water management and conservation efforts. It also relates to the humidity of certain areas and mosquito populations during summer.

\subsubsection{Farmers}
\label{farmers}
Almost all farmers already know the geomorphological characteristics and weather conditions of their farms, thus those services were not very useful to them (see Table \ref{tab2}). They experimented with \textit{tree counting} and \textit{crop classification} services, mainly out of curiosity. For large fields, it makes sense for some farmers to count their trees, or double-check estimated numbers. The analysis of feedback indicates that the agricultural sector prioritizes two key capabilities: advanced warning systems for impending climatic extremes, operating on near-term timescales (hours to days prior to event onset); delivery of tailored, crop-specific advisory services to inform protective actions (see Section \ref{agrowarning}).

\subsubsection{Forestry experts}
\begin{figure}[htbp]
\centerline{\includegraphics[width=0.5\textwidth]{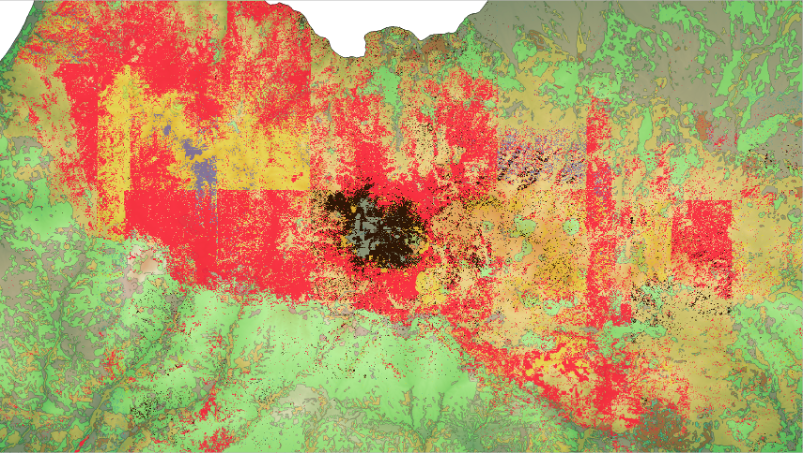}}
\caption{Pinus negra (black) vs. pinus brutia (red) forested areas at Troodos.}
\label{fig2}
\end{figure}

We presented GAEA to experts of the national Forestry Department. Their main aim was to detect the health status of forested areas and have access to early warnings in cases of water stress or pest infestation. The total carbon stock of the island, considering the total biomass of all trees is also well needed, for reasons of reporting and adhering to national/EU policies and goals for carbon emissions. An important issue they are particularly concerned is the introduction of the non-native \textit{Pinus nigra} (Black Pine) to Cyprus Troodos Mountain, which has created a serious ecological threat to the native \textit{Pinus brutia} forest. Pinus nigra aggressively out-competes its native relative, disrupting its natural regeneration cycle. 
The problem is severe, as it leads to the gradual replacement of the endemic ecosystem, reducing biodiversity and fundamentally changing the character of the Troodos forest. Experts wanted to know the \textit{velocity} of spreading of Pinus brutia during the last 10 years, which became possible by exploiting the "tree classification" service of GAEA, described in \cite{amin2024weakly}. Figure \ref{fig2} illustrates the competition between the two tree species as visualized via GAEA.

\section{Lessons learned}
During these three years, having a powerful and innovative tool in our hands, having interacted with various diverse stakeholders and beneficiaries, all this experience allowed us to learn some lessons about the use of environmental EDTs and their usefulness and acceptance by citizens and professionals. We list the key lessons below.

\subsection{Generic vs. specific}
Having a one-fits-all solution as a digital replica of a country is not always useful to certain professionals. Regional authorities and governmental entities want to be able to study "what-if" scenarios and examine their implications on the society, economy and the environment. For example, we were requested to investigate the following scenarios:
\begin{itemize}
    \item Identify the optimal locations to install 1,000 charging stations of electric vehicles, ensuring that every citizen has in average 10 minutes driving time from the nearest station.
    \item Identify suitable locations for photovoltaic parks, respecting regulatory, legal, technical, and practical constraints.
    \item Examine how natural risks at each location are expected to change based on different IPCC climatic scenarios \cite{pirani2024scenarios} in 2030, 2050, and 2100.
    \item Identify the optimal locations to install wildfire extinguisher equipment stations around Cyprus, to ensure that all risk-5 locations (see Table \ref{tab1}) can be reached from those stations within 5-10 minutes.
    \item Study the propagation of a wildfire that was incited at a specific location, based on current climatic conditions and vegetation fuel of the area, creating escape routes for citizens to evacuate safely \cite{karatsiolis2022exploiting}.
\end{itemize}

\subsection{3D modeling and visualizations}
Having a 3D model of a whole country, especially one that is frequently updated, requires massive amounts of data and resource-intensive computing resources \cite{scopigno2017delivering}. While such a feature creates an immediate "wow" reaction, in reality it does not serve much. Only two stakeholder groups discussed about the usefulness of such a feature:
\begin{itemize}
    \item Dept. of Forestry: Estimate the growth rate and total biomass of the carbon stock in forests around the island each month.
    \item Dept. of Town Planning and Housing: All buildings and infrastructures must strictly conform to their designated planning zones in both area and height, ensuring all construction is fully legal.
\end{itemize}
From a research perspective, combining 3D information of an EDT with VR experiences may be useful in educational activities (e.g., teach children about the fauna and flora of the island via images, video and sound from where trees and animals live) \cite{nativi2021digital}, as well as touristic ones (e.g., virtual tours in landmarks of cultural significance) \cite{luther2023digital}. Cost-efficient approaches to map 3D landscapes from single-view satellite images is a great possibility to allow EDTs to embrace such features more easily \cite{karatsiolis2022exploiting}.

\subsection{Use of chatbots and avatars}
\label{Chatbot}
\begin{figure}[htbp]
\centerline{\includegraphics[width=0.5\textwidth]{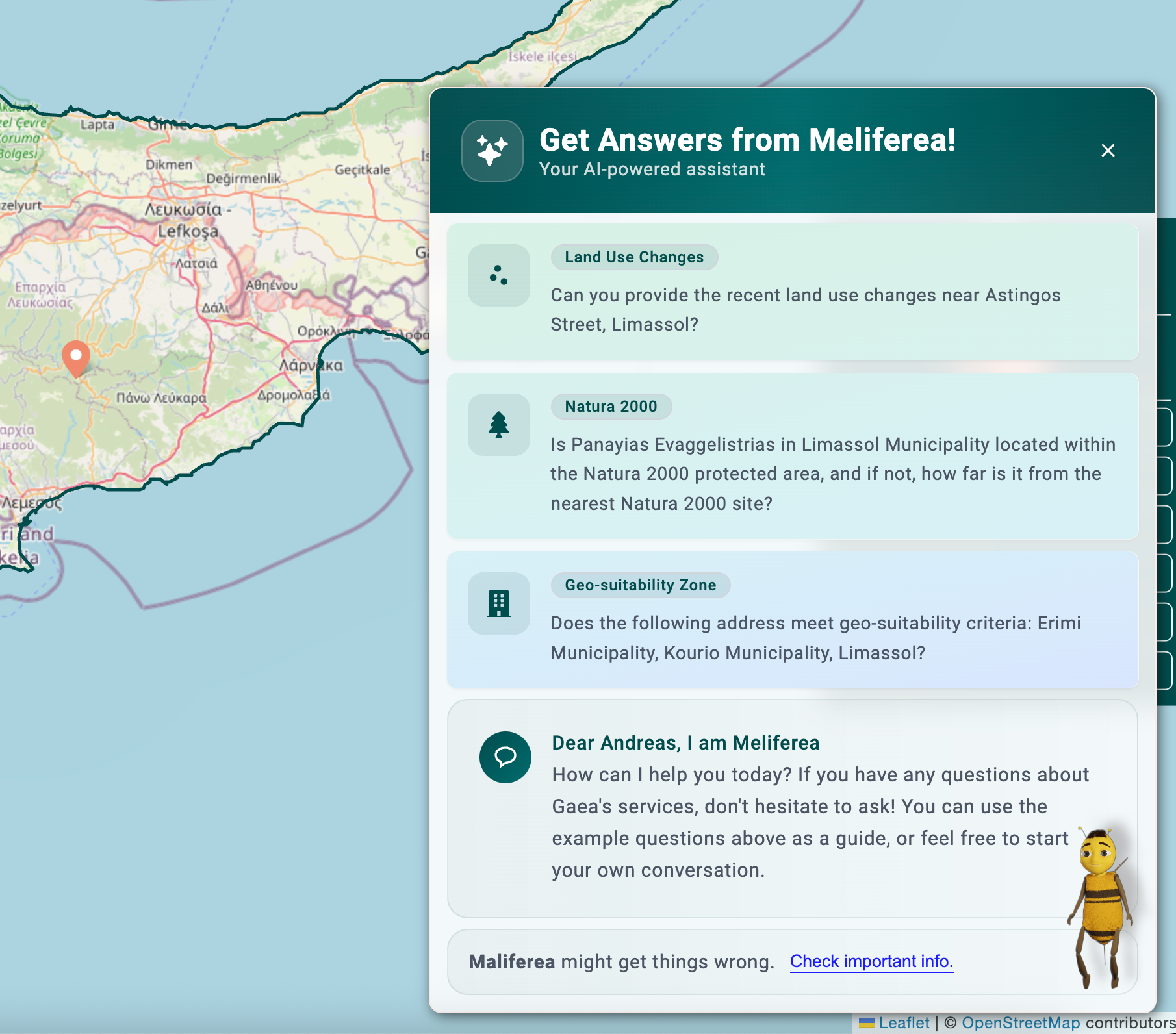}}
\caption{Meliferea avatar on GAEA dashboard.}
\label{fig3}
\end{figure}

Tools such as GAEA, having tens of environmental services, each with its own scope and context, makes it difficult for users to master the tool and harness it properly. After several suggestions from end users, we decided to consider the use of a chatbot, empowered by an LLM \cite{embrechts2025meliferea}, together with an avatar that animates based on the user's predicted emotions while using GAEA \cite{waalAvatar}. The avatar was named \textit{Meliferea} (native bee species of Cyprus) and could answer questions in human language. An example is shown in Figure \ref{fig3}.

In \cite{embrechts2025meliferea}, we examined the possibilities and limitations of LLMs to address human questions of varying complexity related to geospatial analytics. The results from a user-based evaluation indicated good performance in terms
of completely correct answers, with accuracy exceeding 82\%. However, as the complexity of questions increases, involving multiple geoanalytical services and locations together, the system struggles to provide accurate responses and the accuracy rates
drop significantly. Our key finding is that LLMs are capable of transforming the usability of geoanalytical tools and EDTs, lowering the barrier of such tools by beginners or everyday users. However, further research is needed on how to address complex questions related to comparisons (e.g., \textit{Which village is closer to a hospital, A or B?}) or summarized responses (e.g., \textit{From all local communities in region C, which has the largest ratio of trees vs. houses?}). Finally, the use of interactions on the avatar by sensing human emotions helped to slightly increase user engagement with the tool, but our results are not yet statistically significant \cite{waalAvatar}.

\subsection{Modeling and communicating physical risks}
As Table \ref{tab1} indicates, it is very sensitive to share geohazard-related information to the public, and communication language needs to be accurate and specific, not provoking anxiety. Information shared should be transparent, based on facts, as much as possible. The precise methodologies used, performance metrics and scores of the AI models on those metrics need to be clearly communicated to the users, reflecting the reliability of the AI modelling performed. We faced issues communicating the scientific methodology followed to estimate natural risks to banks and insurance companies, who could not understand AI and AI-related performance metrics. We realized that the use of examples and common language works well (e.g., \textit{We predicted 90\% of historic landslide events during the last 10 years}), as well as sensitivity analysis translated into examples (e.g., \textit{The wildfire at location X occurred due to factors Y and Z}). It is worth noting here that the physical risk of land displacement has only recently been added to the list of risks for properties, after certain cases were reported in different parts of the island. We found it very difficult to assign a hazard risk score to land displacement cases, as the literature on this topic is very limited. We eventually adopted a case study from Mexico, which was the most accurate and well modeled \cite{castellazzi2016land}.

\subsection{Different stakeholders, different needs}
\begin{figure}[htbp]
\centerline{\includegraphics[width=0.5\textwidth]{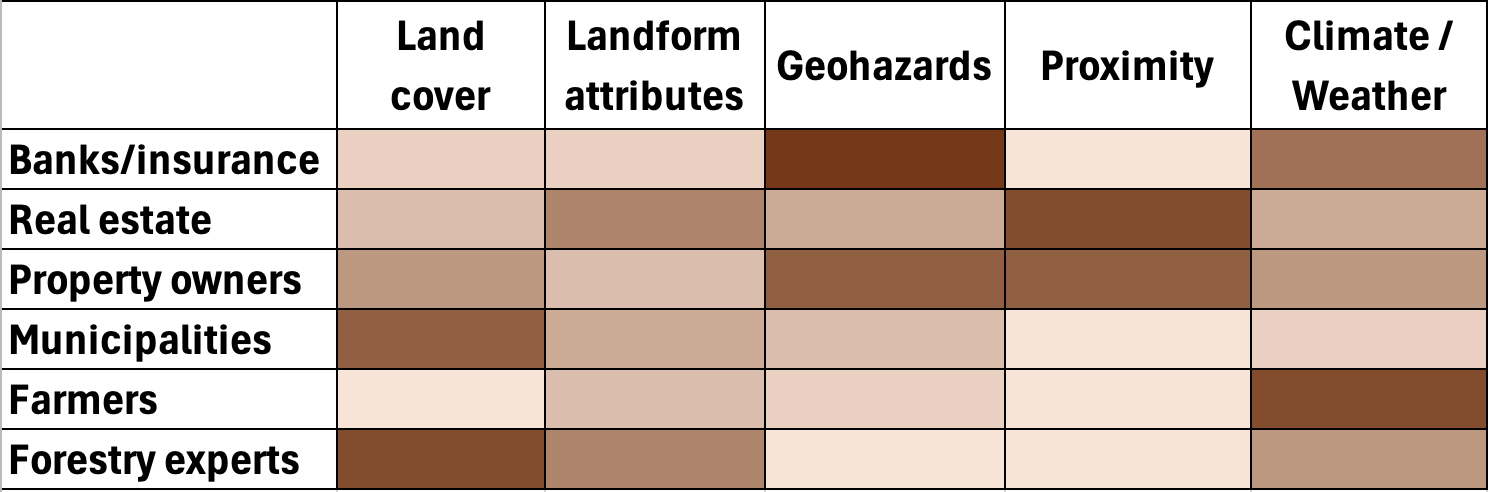}}
\caption{Use of services per stakeholder group. Darker color indicates heavier use. Data was normalized before visualization.}
\label{fig4}
\end{figure}

Each group of stakeholders has different needs, which are translated in the use of different (categories of) services. A summary of the use of service categories vs. stakeholder groups is shown in Figure \ref{fig4}. For example, banks/insurance companies are mostly interested in geohazards, real estate agents in proximity services, property owners in geohazards and proximity, municipalities in land cover services (building areas, swimming pools, trees), the same as forestry experts. Finally, farmers are interested mainly in climatic/weather services and not so much in landform attributes, as they already know about the soils and altitudes of their fields. The main lesson here is that one should design an EDT having in mind the main beneficiaries and their needs, to design relevant services accordingly.

\subsection{Staying up to date}
Keeping data and services up-to-date, when dealing with country-scale EDTs with tens of services is a real challenge, requiring many person-months (PM) per year. For GAEA, the only realistic plan was to fully update its 27 services once a year. As most services are rather static with sporadic changes, this was not a major issue. It would be an issue for other EDTs, where near real-time is needed, e.g., mobility scenarios, traffic estimations, energy load/consumption inside an electricity grid, etc. \cite{mihai2022digital, nativi2021digital}. We estimated that the yearly maintenance of GAEA requires 10-12 PMs, both from full-stack web developers and data scientists/ML engineers. The process needs to start months before the planned yearly update, to send letters to relevant governmental departments and ask for data-related updates (e.g., soil maps, raw data from weather stations, disasters recorded during the previous year etc.) plus updates from online sources (e.g., changes to amenities around the island). The yearly plan of action for updating data and services is summarized in Table \ref{tab2}. An important lesson here is to consider maintenance costs and sustainability before any ambitious design and development of EDTs, especially when updates to data and services need to be frequent. 

\begin{table*}[htbp]
\caption{GAEA yearly plan of updates.}
\begin{center}
\begin{tabular}{|c|p{15.0cm}|}
\hline
\textbf{Service type} & \textbf{\textit{Actions}}  \\
\hline
Land cover & Purchase of high spatial-resolution satellite imagery. Use of AI models taking satellite imagery as input to produce new land cover maps and detection/classification of trees, detection of building areas and swimming pools.\\
\hline
Landform attributes &  Letters to governmental agencies and departments to request for updates to geospatial maps, especially soil mapping.\\
\hline
Geohazards & Recording of new disasters that occurred during the previous year. Re-training of AI models using recent data as additional input, considering also recent updates in landform, land cover and weather/climate information. \\
\hline
Proximity & Updated lists of amenities based on online mapping service providers.  \\
\hline
Climate/weather & Purchase of raw data from 50 weather stations deployed around the island of Cyprus, recording weather information per minute during the last 12 months. Data needs then to be organized hourly, daily, monthly and yearly, before use by the relevant services.  \\
\hline
\end{tabular}
\label{tab2}
\end{center}
\end{table*}

\section{Future challenges}
This section lists current and future challenges for an EDT such as GAEA. We believe that this would be interesting to the reader, as it encompasses common challenges for many EDTs having similar structure, aims and philosophy. We derived those challenges based on own observations, interactions with stakeholders and lessons learned.

\subsection{User-centric design and experience}
Due to the fact that GAEA is a complex geospatial tool, it is important to improve user experience, making things easier for users to navigate through the tool and interact with the services. In this regard, we intend to improve our Meliferea chatbot (see Section \ref{Chatbot}), making it more reliable and accurate. The aim is eventually to be able to understand any human question including geography and geo-analytics, providing proper responses even to complex questions \cite{embrechts2025meliferea}.

\subsection{Optimize Validation and Testing}
Right now, validation occurs by randomly selecting tens of scenarios for every service, which requires hundreds of tests every time there is a major update, while hundreds of other scenarios escape testing. Validations need to become more automatic, reducing the human intervention and inspection, plus considering testing scenarios more wisely. Unsupervised clustering is a promising technique for mapping data and scenarios together into geospatial clusters \cite{sadeghi2025clustering}, testing only \textit{representatives} of each cluster.

\subsection{AI-Based Risk Modeling}
The goal here is to accelerate the process needed to calculate the risks of natural hazards all around the island of Cyprus, considering faster updates of the features needed to calculate risks, plus better and faster incorporation of recent disasters (used as ground truth), to better train and test the models involved. Automatic pipelines for feature engineering are much needed, especially in the case of wildfires.

\subsection{Incorporating climate projections and scenarios}
Right now, GAEA  does not consider climate projections for natural risks. At the same time, certain risks (i.e., coastal and flash flooding, wildfires, landslides), are susceptible to climatic changes. We aim to incorporate IPCC projections on climate changes by 2050-2100 \cite{pirani2024scenarios}, and include those projections to the re-calculations of the risks, allowing users to experiment with different climatic scenarios and their impact on natural disasters.

\subsection{Disaster response scenarios}
Our natural hazard risk models do an excellent job to characterize the hazard level of any location on the island of Cyprus, from 1 (no risk) to 5 (very high risk). However, these hazard maps have not been harnessed to derive new disaster response plans or to align with existing ones. We want here to perform both, for example, to align flooding with the existing plans of certain municipalities of Cyprus (i.e., Nicosia, Limassol), plus consider disaster response plans such as \textit{“citizens evacuating high-risk areas in case of wildfire”} or \textit{“map the propagation of an on-going wildfire”}, based on our existing vegetation and land cover maps \cite{kamilaris2023examining}. 

\subsection{Adapting for real estate}
The reader may observed that GAEA’s structure is suitable for the real estate market (see Section \ref{realestate}). Our services are appropriate for real estate managers and agents, and property owners, who may better inspect and valorize properties, incorporating natural risks in the inspection process. We intend to better investigate the impact of geospatial insights from EDTs on the performance of AVMs \cite{gravier2024leveraging}, enhancing basic property characteristics together with geospatial analytics (natural risks, distance to amenities or points of interest, location-based socioeconomic indicators, etc.).
Another interesting potential of EDT-empowered geospatial intelligence is to create more personalized, targeted and intelligent recommendation systems for properties, where potential buyers can find the best property tailored to their unique needs \cite{yuan2013toward}.

\subsection{What-if scenarios}
An important challenge, requested again and again by policy-makers and governmental departments, is to run simulations and \textit{what-if} scenarios, examining the impact of various actions (e.g., capacity for further installation of photovoltaics parks, best locations for afforestation actions, strategic placement of electric vehicle charging stations, etc.). We aim to extend GAEA to allow such simulations and scenarios to be executed correctly and accurately, allowing policymakers to investigate the potential impact of their decisions on the society, economy, and the environment.

\subsection{Biodiversity monitoring}
We received valid feedback from researchers and the Department of Environment regarding the exclusion of wildlife and biodiversity data from our EDT. This omission stemmed from a lack of accurate, curated and complete datasets on the island's fauna. Key priorities include rodent populations, for health and hygiene, and pollinators, crucial for agriculture and food security. For both rodents \cite{padubidri2025spyce} and pollinators \cite{padubidri2023hive}, we developed AI-assisted camera traps to estimate populations. This task is complex due to challenges in covering large, diverse landscapes and accounting for seasonal changes \cite{greco2025placement}. Seasonality is critical for biodiversity, yet GAEA's current structure supports only annual updates.

\subsection{Agro-meteorological warning system}
\label{agrowarning}
Farmers and the Departments of Agriculture and Meteorology (see also Section \ref{farmers}) are mostly interested in an integrated agro-meteorological early warning and decision support system for extreme weather events, such as frost, heatwaves, heavy rainfall, and hail. The development of an intelligent AI/ML system capable of detecting hazardous phenomena early and providing immediate and targeted warnings to relevant authorities and the public will significantly improve preparedness and crisis management \cite{cova2017warning}. Current forecasting services do not fully integrate local specifics nor provide targeted and timely warnings for extreme weather events that directly affect the agricultural sector. In this regard, we plan to develop an AI system that integrates real-time data and forecasts on GAEA, to produce highly accurate, personalized warnings for farmers. Notifications will be delivered through user-friendly channels (mobile app, SMS, etc.), enhancing the resilience and efficiency of agricultural production. The warning system is expected to offer multiple benefits: improving water resource management, protecting crops, reducing financial losses, and optimizing decision-making for farmers. This warning system could then expand to other critical sectors, such as public health, energy, and transportation.

\section{Conclusion}
In this paper, we described the experiences and lessons learned after the deployment of a country-scale environmental digital twin on the island of Cyprus for three years. This digital twin, called GAEA, contained 27 environmental geospatial services and was used by a diverse range of stakeholders with different needs. The paper discussed lessons learned, current challenges of environmental digital twins and future work.


\bibliographystyle{IEEEtranN}
\bibliography{references}

\vspace{12pt}

\end{document}